# ORION: Unifying Top-Down and Bottom-Up Chemical Space Sampling for a Universal Organic Force Field

Zherui Chen [a, b], Jiayu Zhang [c], Yuxuan Tian [d*], Zhoulin Liu [e], Sining Dai [f], Yanghui Li [f], Cong Chen [f], , Dingyuan Tang [a], Yajun Deng [a*], Qingxia Liu [a, g*]

[a] Future Technology School, Shenzhen Technology University, Shenzhen 518118, P. R. China

[b] College of Applied Sciences, Shenzhen University, Shenzhen 518060, P. R. China

[c] Heilongjiang Provincial Key Laboratory of Oil & Gas Chemical Technology, Northeast Petroleum University, Daqing 163318, P. R. China

[d] Beijing Key Laboratory of Green Chemical Reaction Engineering and Technology, Department of Chemical Engineering, Tsinghua University, Beijing 100084, P. R. China

[e] School of Science, Harbin Institute of Technology, Shenzhen 518055, PR China

[f] School of Energy and Power Engineering, Dalian University of Technology, Dalian 116024, P. R. China

[g] Department of Chemical and Materials Engineering, University of Alberta, Edmonton, AB T6G 1H9, Canada

*Corresponding author

**Abstract:** Empirical force fields remain the primary tool for large-scale molecular simulation, yet their limited flexibility and transferability often hinder predictive modeling in chemically complex condensed-phase systems. Here we present ORION, a universal machine-learning force field for C, H, O, N, S, and P systems developed within the Neuroevolution Potential (NEP) framework. To enhance transferability across diverse chemical environments, ORION was trained on a chemically rich dataset constructed through an integrated top-down and bottom-up strategy, enabling accurate descriptions of complex organic configurations, reactive intermediates, and weak intermolecular interactions. ORION achieves near-density-functional-theory accuracy while retaining the efficiency required for large-scale molecular dynamics simulations. On the test set, it predicts atomic forces with substantially higher accuracy than ReaxFF while running 215.5 times faster under identical hardware conditions, making simulations on the hundreds-of-nanoseconds timescale readily accessible. The model provides a balanced description of bond breaking and formation, aromatic growth, hydrogen bonding, van der Waals interactions, and π-stacking, demonstrating strong transferability across both reactive and nonreactive systems. These results establish ORION as a practical and general force field for predictive simulations in chemistry and materials science, and provide an effective route toward universal machine-learning force fields with both high accuracy and broad applicability.

# 1 Introduction

Over the past few decades, molecular simulations have become indispensable tools for understanding microscopic interactions that are challenging to probe experimentally. Classical molecular dynamics (cMD), based on Newtonian mechanics, utilizes predefined force field potential surfaces to compute atomic forces efficiently[1-3]. However, the rigid parametrization of conventional non-reactive force fields limits their ability to capture dynamic chemical bond evolution, rendering them unsuitable for modeling complex chemical reactions. Reactive force fields such as ReaxFF have advanced the simulation of chemical reactivity but remain constrained by their reliance on empirical parameterization tailored to specific chemical environments[4-6]. These limitations include difficulties in describing bond order transitions, energy changes, and multi-atomic cooperative effects during reaction dynamics. Additionally, ReaxFF requires extensive manual optimization, limiting its generalizability across chemically diverse systems. In contrast, ab initio molecular dynamics (AIMD) offers unparalleled accuracy and flexibility by directly computing forces from quantum mechanical principles, making it well-suited for modeling chemical reactions[7-10]. However, AIMD is computationally expensive, limiting its applicability to large-scale systems or long timescales. The trade-off between computational efficiency and accuracy highlights the urgent need for a reactive potential that is fast, accurate, and universal, capable of reliably predicting reaction mechanisms and simulating entire processes prior to experimental validation.

Machine learning (ML) has recently transformed molecular dynamics by offering a novel approach that combines computational efficiency with high accuracy. By learning high-dimensional representations of atomic environments from quantum mechanical reference data, machine learning potentials (MLPs) achieve AIMD-level accuracy while approaching the computational speed of classical force fields[11-14]. This dual advantage opens the door to modeling complex chemical reactivity with unprecedented efficiency. Zhang et al. [15] developed an ML potential for ZrB$_2$ that accurately predicts structural, elastic, and phonon properties across temperatures from room temperature to 2500 K, including thermal expansion and transport characteristics. Similarly, Berger et al. [16] employed ML potentials to model phase transitions in perovskites and predict polarization changes during these transformations. While MLPs have achieved remarkable success in inorganic and metallic systems, their application to organic reactions remains underexplored and faces unique challenges. Organic molecules exhibit exceptional chemical diversity, encompassing a wide range of bonding patterns, functional group combinations, and molecular conformations. Accurate modeling of weak intermolecular forces—such as hydrogen bonding, van der Waals interactions, and π–π stacking—is critical for capturing the reaction dynamics of organic systems[17-19]. These weak interactions often play key roles in determining reaction pathways, transition states, and kinetics. Furthermore, the vast chemical space of organic compounds complicates the development of comprehensive, high-quality training datasets, which are essential for constructing MLPs that can generalize across diverse organic systems and reaction mechanisms. Addressing these challenges is crucial for unlocking the full potential of MLPs in organic chemistry.

Current machine learning potentials for organic systems, such as ANI-1x[20], MACE-OFF[21], provide a starting point but remain limited in scope. These models primarily focus on CHON elements and rely on "bottom-up" approaches that generate molecular geometries through random atomic combinations. While this strategy facilitates rapid dataset generation, it often lacks representation of realistic chemical systems and fails to capture the complex interplay between functional groups, molecular architectures, and reactive intermediates. As a result, these models may struggle to predict intricate reaction pathways or dynamics in practical settings, underscoring the need for new approaches that accurately capture the chemical complexity of organic systems.

Here we develop ORION an **O**rganic **R**eactive Interat**O**mic **N**euroevolution potential[22, 23], enabling accurate simulations of reactive dynamics in systems comprising C, H, O, N, S, and P elements. The training dataset was constructed using a combined top-down and bottom-up approach, designed to capture both the complexity of realistic macromolecular systems and the fundamental chemistry of smaller reactive fragments. This dual strategy ensures broad coverage of chemical space, encompassing diverse bonding environments, functional groups, and weak non-bonded interactions such as hydrogen bonding, van der Waals forces, and π–π stacking. Trained on this chemically diverse and balanced dataset, ORION demonstrates high fidelity in modeling bond rearrangements, transition states, and reaction intermediates, enabling precise simulations of complex reaction pathways across a wide range of organic systems. By delivering near-DFT accuracy with orders-of-magnitude computational speed-ups, ORION bridges the gap between efficiency and accuracy in reactive molecular simulations. Its robustness and transferability make it a powerful tool for elucidating reaction mechanisms and predicting complex chemical networks, with applications spanning energy conversion, catalysis, polymer science, and biochemistry, and it lays a strong foundation for the rational design of advanced materials and processes.

## 2 Reference Data Generation

### 2.1 Dataset Acquisition

The dataset generation workflow is illustrated in Figure 1. The dataset was assembled through an integrated top-down and bottom-up strategy, applied across all configuration types to maximize chemical diversity within CHONSP element space. All training structures were stored in the extended trajectory file format (etxyz) with a config_type label used to distinguish their origins. In the top-down direction, reactive configurations were sampled from 3000 K semi-empirical molecular dynamics (GFN1-xTB, CP2K 2024.1, NVT ensemble) of complex macromolecular systems such as coal, asphaltenes, proteins, carbohydrates, and nucleic acids, together with their associated reactions involving small molecules and mixtures. In the bottom-up direction, representative small-molecule structures—including those from PubChem[24]—were systematically perturbed via dihedral scans and coordinate displacements, and combined or extended to construct larger molecular frameworks. Additional datasets from existing literature were curated to further enrich functional group diversity and reaction coverage. This unified dual-pathway workflow, spanning from complex condensed-phase systems to fundamental reactive fragments, yields a chemically rich and systematically balanced dataset for training a transferable machine-learning reactive potential. The composition of all datasets is shown in Table 1. The training set contains 68579 structures with a total of 10634014 atoms.

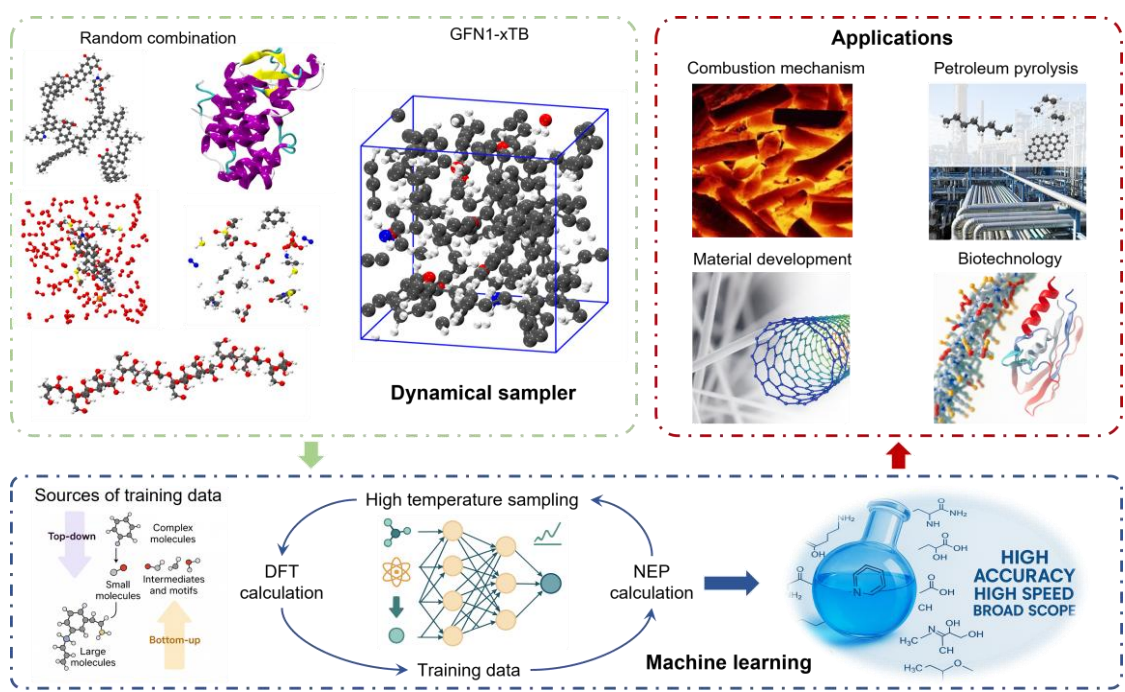

Figure 1. Data acquisition process

Table 1. Dataset composition

| config_type | data source | calculation method |
|---|---|---|
| cp2k2xyz | this study | PBE-D3/DZVP-MOLOPT-SR-GTH (CP2K) |
| orca2xyz | this study | ωB97M-V/def2-QZVP (ORCA) |
| ANI-1xnr | reference[20] | BLYP-D3/TZV2P-MOLOPT-SR-GTH (CP2K) |
| C2H5O2N | reference[25] | M06-2X/QZV3P-MOLOPT-SR-GTH (CP2K) |
| CCSDT-water | reference[26] | CCSD(T)/aug-cc-pVTZ (CFOUR) |
| CO2 | reference[27] | SCAN-rVV10 (Quantum-ESPRESSO) |
| GO | reference[28] | PBE-D3 (CASTEP) |
| graphene-water | reference[29] | optB88-vdW/TZV2P-MOLOPT-SR-GTH (CP2K) |
| Graphene_N_Pyridine | reference[30] | PBE-MBD-NL/tight (FHI-AIMS) |
| MACE | reference[21] | ωB97M-D3(BJ)/def2-TZVPPD (PSI4) |
| MB-pol | reference[26] | MB-pol (lammps) |
| NH3-CO2 | reference[31] | SCAN (Quantum-ESPRESSO) |
| QCArchive | reference[32] | B3LYP-D3(BJ)/def2-SVP (PSI4) |
| Transition1x | reference[33] | ωB97x/6–31 G(d) (ORCA) |
| Liquid-Electrolyte | reference[34] | ωB97x-D3(BJ)/def2-TZVPD (PSI4) |
| md22 | reference[35] | PBE-MBD/tight (FHI-AIMS) |
| oil-water | reference[36] | M06-2X/QZV3P-MOLOPT-SR-GTH (CP2K) |
| omol25 | reference[37] | ωB97M-V/def2-TZVPD (ORCA) |
| protein | reference[38] | PBE0-MBD/def2-TZVPP (PSI4) |
| spice | reference[39] | ωB97M-D3(BJ)/def2-TZVPPD (PSI4) |

**2.2 Training Process**

ORION was developed following the standardized Neuroevolution Potential (NEP) training protocol, which requires three primary input files: the training dataset (train.xyz), the testing dataset (test.xyz), and the NEP input parameter file (nep.in). All NEP training and validation procedures were performed using the GPUMD package, a high-performance open-source platform for machine-learning interatomic potentials[22, 23]. To ensure a consistent energy baseline across datasets originating from different electronic-structure codes, we first performed energy shifting for the subset with config_type = cp2k2xyz. In this step, atomic reference energies were optimized to minimize the absolute magnitude of the total energies, yielding values of C: −153.416775 eV, H: −16.447581 eV, O: −434.675726 eV, N: −268.508677 eV, S: −261.553707 eV, and P: −207.157188 eV. The shifted cp2k2xyz data were then used for an initial NEP training, producing a preliminary force-field file (ORION0.txt). Subsequently, all remaining DFT datasets were aligned to the ORION0.txt baseline using an evolutionary optimization algorithm that determines energy shifts by minimizing the mean-squared error with respect to the preliminary model. This approach ensures that total energies from heterogeneous sources are placed on a unified reference scale while preserving physically meaningful energy differences. Following this alignment, the combined dataset was used for continued NEP training, resulting in the final ORION potential. The energy-shifting procedure is formulated in Eq. (1) and has been implemented as part of the NepTrainKit software suite[40].

$$E_i^{\text{shift}} = E_i^{\text{DFT}} - \sum n_{i\text{X}} \Delta(E_\text{X}^{\text{target}}) \qquad (1)$$

where $n_{i\text{X}}$ is the number of atoms of element X in the configuration and $E_\text{X}^{\text{target}}$ is the corresponding reference single-atom energy. Since only relative energy differences are physically relevant for force-field training, this transformation does not affect the accuracy or transferability of the resulting potential.

To assess model performance, the ORION-predicted energies and forces were directly compared to the corresponding DFT reference values, as shown in Figure 2. The root mean square errors (RMSEs) for both the training and testing datasets are reported, demonstrating consistently low values across all target properties.

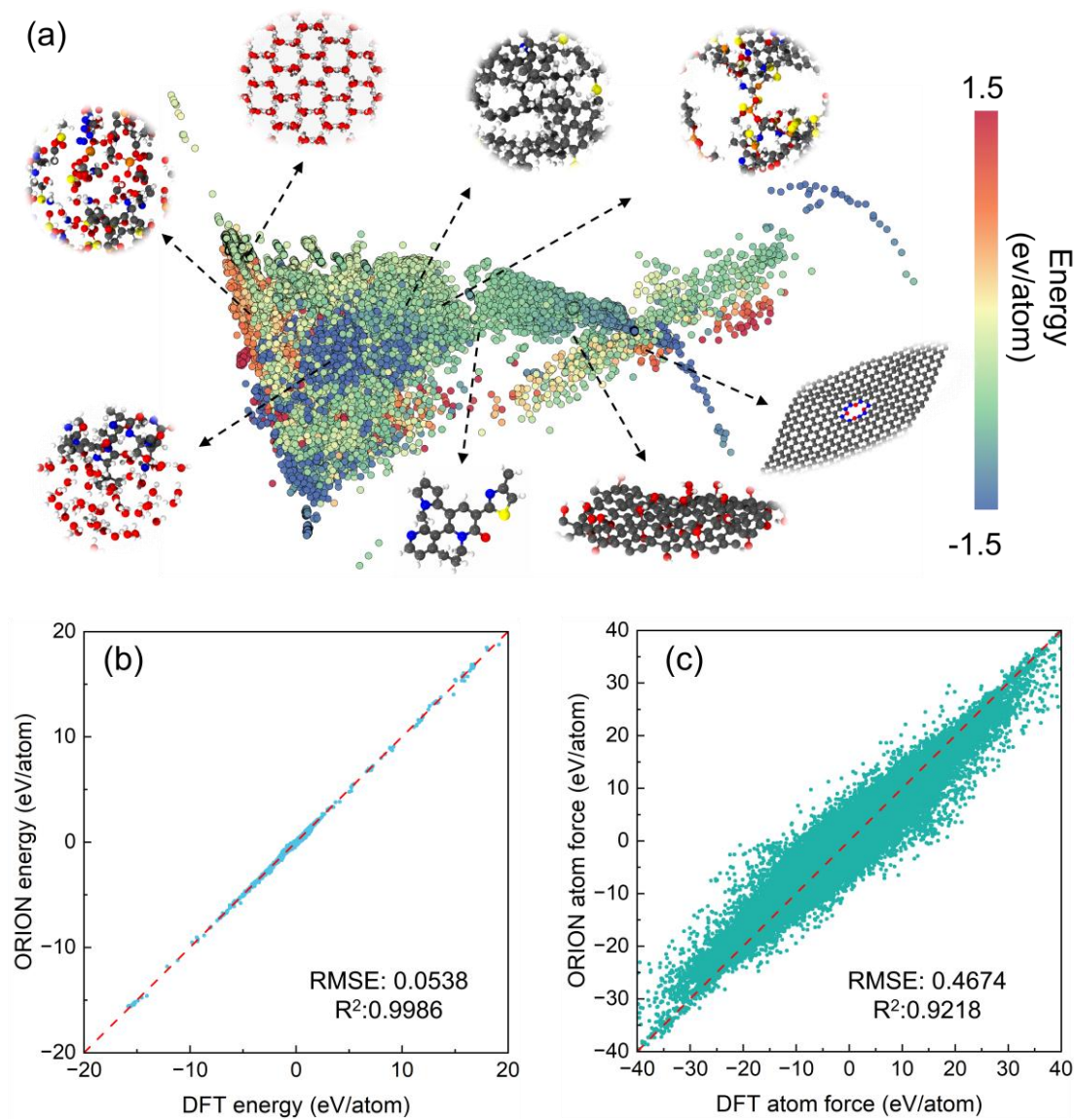

Figure 2. (a) PCA dimensionality reduction distribution characteristics of training set；(b) Comparison of DFT energy and ORION energy in the training set; (c) Comparison of atomic forces between DFT and ORION in the training set

For benchmarking, we compared the ORION directly with the ReaxFF reactive force field[41, 42], as our primary focus is on accurately modeling reactive processes. The comparison was carried out on the test.xyz dataset, we evaluated only atomic force predictions as shown in Figure 3. The force RMSE for ReaxFF was substantially higher than that of the NEP model, highlighting the superior accuracy of our machine learning-based potential in reproducing DFT-calculated atomic forces. In addition to its accuracy, the ORION demonstrates remarkable computational efficiency: when both methods are executed on an NVIDIA RTX 4090 GPU, the ORION achieves a speedup of 215.5 times compared to ReaxFF. This result underscores that, while maintaining DFT-level accuracy, the ORION also enables ultrafast simulations, offering both reliability and high-throughput capability for studying reactive dynamics in complex organic systems.

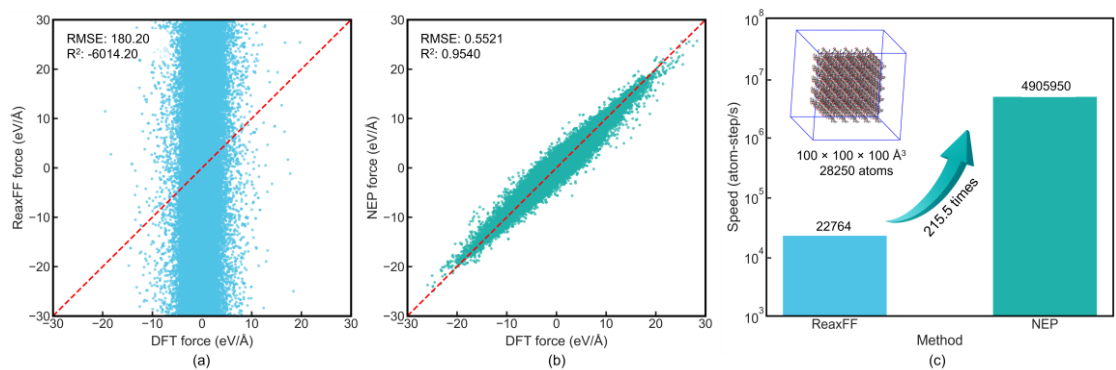

Figure 3. Performance of ORION and ReaxFF force fields in the test.xyz. (a) Atomic forces calculated by ReaxFF; (b) ORION in test.xyz; (c) Comparison of calculation speed

## 3 Results and Discussion

### 3.1 Microscopic Processes of Combustion

Combustion involves complex reaction networks and multiscale mechanisms, posing challenges for conventional approaches in resolving key intermediates and transient species. To address this, we employ the ORION to investigate atomistic combustion mechanisms. Using the methane combustion model from ANI-1xnr[20] as a benchmark, ORION accurately reproduces the major products (Figure 4), demonstrating its reliability for simulating fundamental combustion reactions.

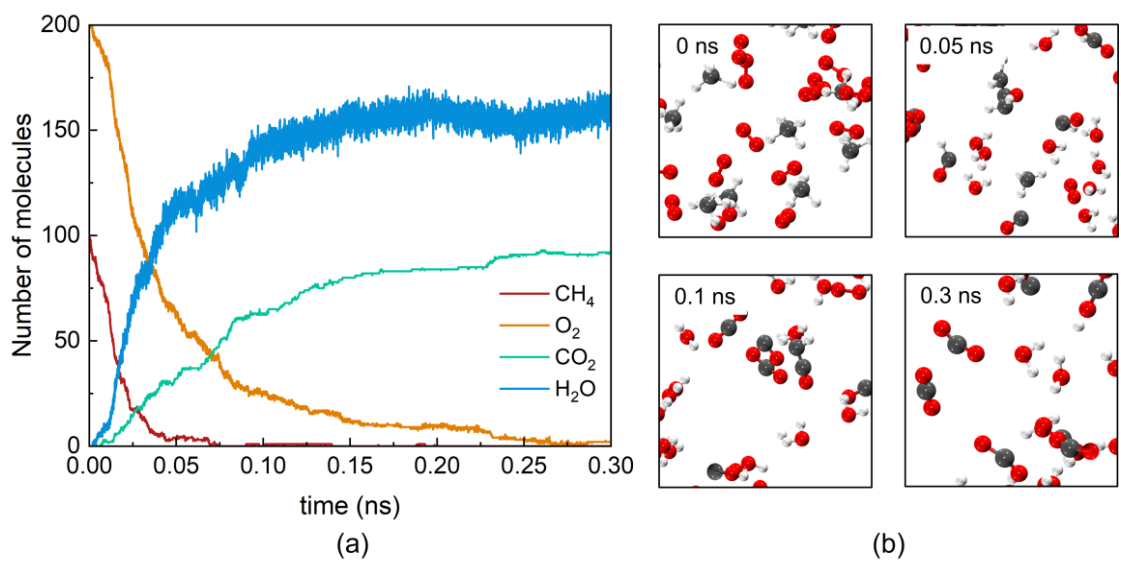

Figure 4. Methane combustion reaction. (a) Number of main reactants and product molecules; (b) Reaction snapshot.

Coal combustion presents a far greater mechanistic challenge than methane oxidation due to its intrinsically heterogeneous molecular architecture and the presence of diverse functional groups, cross-linking motifs, and heteroatoms. Understanding its atomistic reaction pathways remains difficult yet critical for advancing clean-combustion technologies and elucidating pollutant-formation mechanisms. To extend the applicability of ORION to such complex systems, we investigated the combustion of lignite as a representative case. A model consisting of ten lignite molecules ($C_{225}H_{182}O_{36}N_4S_3$)[43] was constructed and combined with 500, 2500, or 5000 $O_2$ molecules to represent different oxygen-availability conditions. After non-reactive equilibration using GROMACS with the GAFF force field, reactive simulations were performed with ORION in the NVT ensemble at 2000 K for 10 ns using a timestep of 0.5 fs.

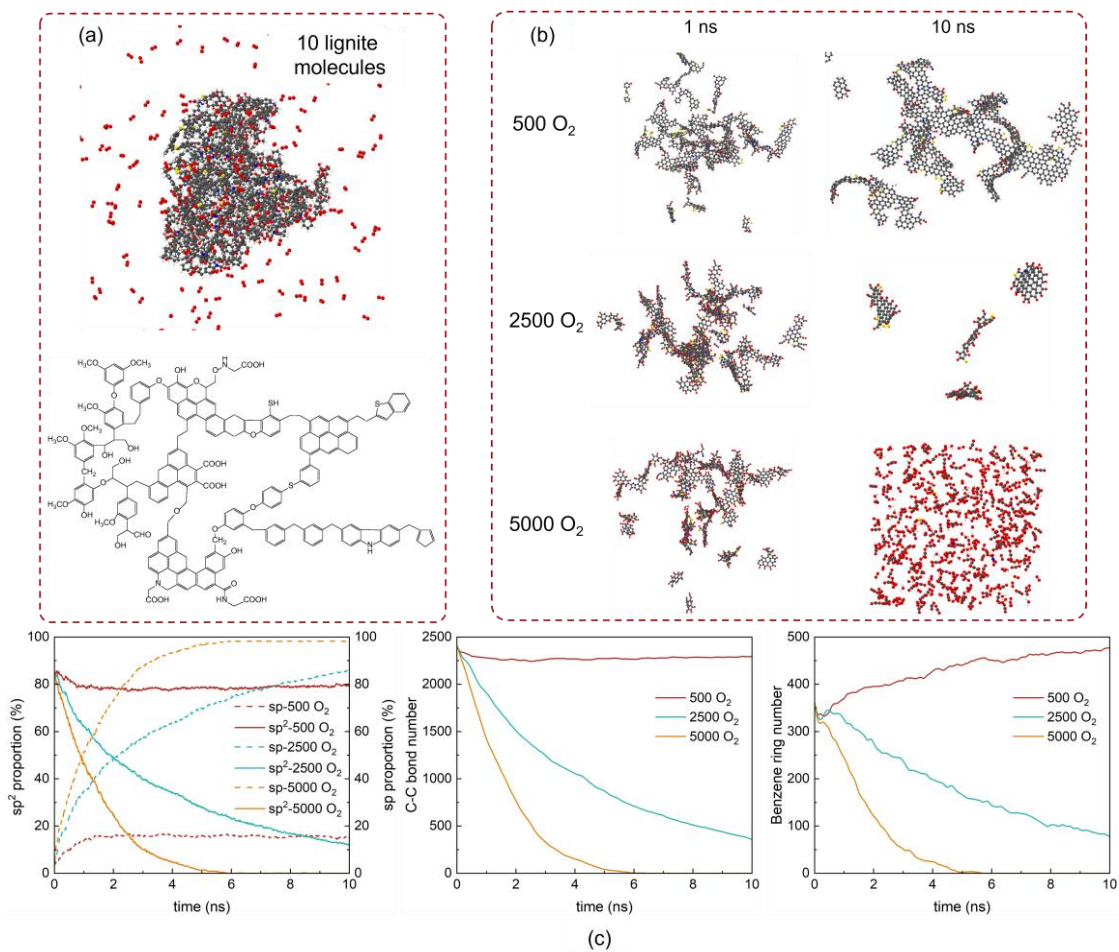

Figure 5. Combustion process of lignite under different oxygen concentrations. (a) Initial model and molecular structure of lignite; (b) Combustion reaction snapshot; (c) Dynamic changes in molecular structure during combustion process. For clarity, only the organic components are shown in the figure for the cases of 500 and 2500 $O_2$ molecules.

The structural evolution and reaction pathways during the oxidation of brown coal are highly dependent on oxygen availability, as illustrated in Figure 5. When the number of oxygen molecules increases from 500 to 5000, the dominant reaction mechanism shifts progressively from condensation and aromatization toward complete oxidation. Under oxygen-deficient conditions (500 $O_2$), the macromolecular network of lignite undergoes thermal cleavage in the initial pyrolysis stage, generating numerous radical fragments. However, due to the limited concentration of oxygen radicals, these fragments are not effectively oxidized; instead, they undergo recombination and aromatization, forming more stable polycyclic aromatic structures. This process is reflected in a

31.77% increase in the number of benzene rings within 10 ns, while the number of C–C bonds stabilizes after a transient decrease. The persistently high proportion of $sp^2$-hybridized carbon atoms indicates the formation of coke-like precursors. When the oxygen content is raised to 2500 molecules, oxidative ring-opening and bond-breaking reactions become dominant. The organic framework undergoes continuous depolymerization, and the final $sp^2$ carbon content decreases to 11.78%, demonstrating that the rate of oxidative fragmentation surpasses that of radical recombination, shifting the reaction pathway from aromatization toward gradual carbon skeleton degradation. In an oxygen-rich environment (5000 $O_2$), lignite undergoes rapid and complete combustion. Aromatic structures collapse entirely within 5 ns, and almost no organic molecules remain after 6 ns. Notably, the proportion of sp-hybridized carbon approaches 100% by the end of the reaction, confirming the complete conversion of organic carbon into gaseous products such as CO and $CO_2$. Kinetically, oxygen-rich combustion exhibits two distinct stages: the initial thermal cleavage of weak bridges (e.g., C–O and aliphatic C–C bonds) generates organic radical fragments, followed by the rapid capture and oxidation of these fragments by the abundant O and OH radicals in the environment, which effectively suppresses secondary condensation and drives the system toward complete oxidation. These results demonstrate that the ORION force field provides reliability, and predictive capability in simulating complex reaction pathways within carbon-based systems like lignite.

**3.2 Development of Carbon Materials**

Bottom-up synthesis of carbon materials can be achieved via alkane pyrolysis, where thermal decomposition initiates dehydrogenation, cyclization, and condensation, leading to carbonaceous networks and graphite-like domains. Resolving the underlying bond-breaking and bond-forming events at the molecular level is essential for linking processing conditions to carbon yield and microstructural evolution. Therefore, we employ ORION molecular dynamics simulations to interrogate these elementary steps during n-octane pyrolysis, providing mechanistic guidance for controlling structure evolution in carbon-material synthesis. The model contains a total of 100 n-octane molecules, with a pyrolysis temperature of 2000K. The results are shown in the Figure 6.

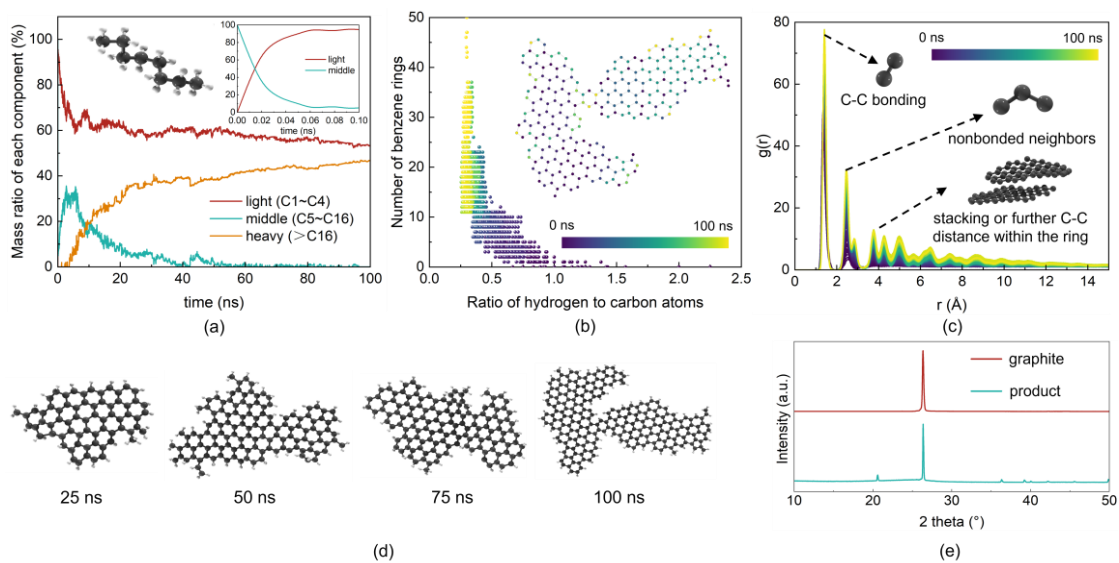

Figure 6. Thermal decomposition-recombination process of n-heptane. (a) Proportion of mass fractions of each component in the reaction production; (b) Aromatization evolution within the largest carbon cluster; (c) The radial distribution function (RDF) of C-C atoms during the reaction process; (d) The structure of the largest aromatic hydrocarbon at each moment; (e) Comparison of XRD patterns between the final solid product and graphite.

For clarity, species with carbon numbers C1–C4 are categorized as light fractions, C5–C16 as middle fractions, and those above C16 as heavy fractions. As shown in Figure 6(a), n-octane is almost completely converted into light species within 0.1 ns, indicating that rapid C–C bond scission and small-molecule formation dominate the initial stage. The subsequent sharp decrease in the light fraction suggests that these small fragments act as reactive intermediates and are progressively consumed via secondary reactions, giving rise to a canonical bottom-up carbonization pathway of "cracking–recombination–condensation/incipient graphitization". The middle fraction reaches a maximum at approximately 5 ns and is already dominated by aromatic hydrocarbons at this point, implying substantial dehydrogenation, cyclization, and aromatization and marking a critical transitional stage toward condensed carbon frameworks. Thereafter, both light and middle fractions continue to decline, whereas the heavy fraction (>C16) increases monotonically and becomes dominant at late times, demonstrating that radical addition, polymerization/cross-linking, and aromatic ring fusion collectively drive continuous backbone growth and conversion into solid-like carbonaceous clusters. Ultimately, the heavy fraction accounts for 46.86% of the products, while the remainder consists primarily of gaseous light hydrocarbons, indicating pronounced solid carbon formation under the present conditions albeit with non-negligible volatile products.

Figure 6(b) provides a direct structural metric for the aromatization process within the largest carbon cluster. With reaction progress, the overall H/C ratio decreases markedly while the ring number increases, showing that light/middle intermediates undergo sustained H abstraction to form aromatic radicals, followed by addition, cyclization, and fusion reactions preferentially occurring at aromatic radical and edge sites. The coupled evolution of decreasing H/C and increasing ring counts thus serves as a quantitative signature of deepening condensation, reflecting an irreversible transition from hydrogen-rich aliphatic structures to hydrogen-poor aromatic/carbonaceous frameworks.

The RDF analysis further supports this structural evolution across length scales. In the C–C RDF, the first peak at ~1.5 Å corresponds to bonded nearest neighbors. As the reaction proceeds, the progressive conversion of $sp^3$ carbon to $sp^2$ carbon and the increasing prevalence of conjugated/aromatic bonding enhance local structural order, leading to a strengthened and more structured first-neighbor peak. In the medium-range regime (3–6 Å), a series of increasingly pronounced oscillatory peaks emerges over time. These features mainly originate from correlations among more distant C–C pairs within and between stacked aromatic fragments/incipient graphitic layers, indicating that the system evolves from flexible chains and small clusters toward a more rigid sp$^2$ network with discernible medium-range ordering (graphite-like short-range stacking). The final H/C ratio within the largest carbon cluster is only 0.29, while there are over 50 benzene rings, exhibiting highly aromatic condensation characteristics.

To validate the simulation results, n-octane was pyrolyzed experimentally at 870 °C and the solid residue was characterized by XRD. The dominant diffraction peak appears at ~26°, close to the graphite (002) reflection, confirming the formation of graphitic interlayer stacking. Meanwhile, the broadened peak shape and the presence of minor additional reflections suggest limited crystallite size and appreciable disorder due to finite reaction time and incomplete structural maturation, in contrast to highly crystalline graphite. Overall, the XRD observations are consistent with the ORION predictions regarding aromatization-driven condensation and the emergence of graphite-like stacking.

Besides graphite-based materials, carbon nanotubes (CNTs) are important carbon allotropes with broad applications in gas separation, electronics, and composite materials. Their performance, however, strongly depends on achieving well-dispersed, ideally individualized CNTs[44-46], which remains difficult because strong van der Waals attraction and π–π stacking promote bundling. To evaluate whether ORION can be used for rapid solvent screening in CNT dispersion, we performed molecular dynamics simulations on a model system containing two CNTs (length 30 Å, diameter 10 Å) in three representative solvents: benzyl alcohol, benzene, and methanol. The total calculation time is 10 ns, and then the last 1 ns is taken to analyze the data. The dispersion behavior was quantified by the center-of-mass distance between the two CNTs, as shown in Figure 7.

Using ORION, benzyl alcohol gives the largest CNT–CNT separation (29.38 Å), indicating the best dispersing ability among the three solvents. This trend is consistent with our experimental observations, where SEM images show a noticeable presence of individualized CNTs in benzyl alcohol. The superior performance of benzyl alcohol can be attributed to its ability to simultaneously provide stabilizing π–hydrogen bond and π–π interactions with CNT surfaces, whereas methanol mainly contributes π–hydrogen bond interactions and benzene primarily provides π–π interactions. By contrast, GAFF predicts only minor differences in CNT separation among the three solvents, indicating that it underestimates the coupled effects of π-hydrogen bonding and π–π interactions that govern solvent-dependent CNT dispersion. These results demonstrate that ORION can more reliably capture solvent-dependent CNT dispersion and is therefore promising for rapid screening of suitable dispersants.

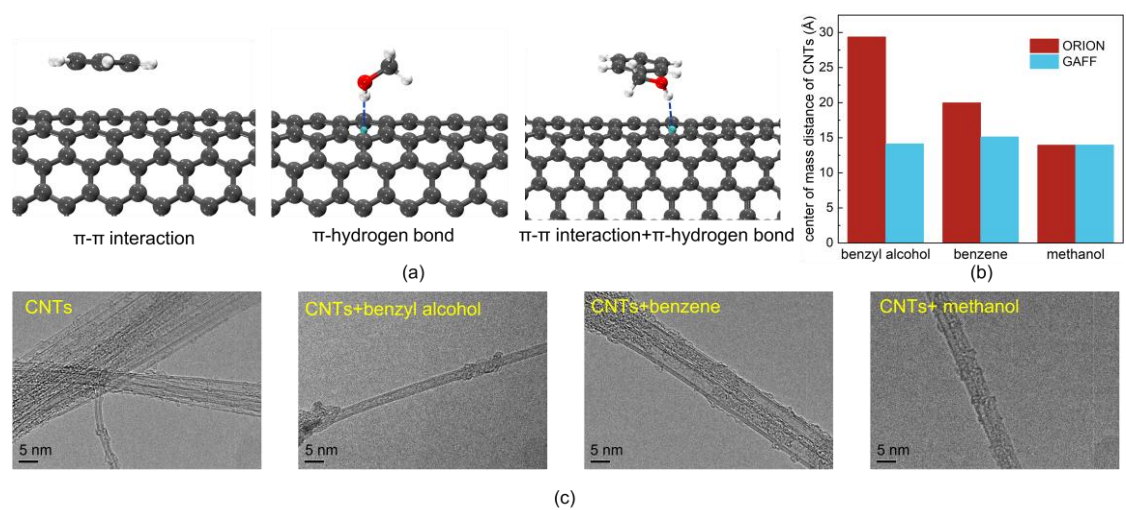

Figure 7. CNT separation process. (a) The interaction between CNTs and solvents; (b) Distance between centroids of CNTs; (c) SEM image of the separated experiment.

### 3.3 Supramolecular and Host-guest Interaction

Molecular crystals are an important target for force-field development because their stability is governed by a delicate interplay among crystal packing, directional intermolecular interactions, and thermally activated molecular motion. As a representative example, we consider sI methane clathrate hydrate, in which a hydrogen-bonded water framework encloses methane molecules within polyhedral cages as shown in Figure 8(a). This system simultaneously probes host hydrogen bonding, host-guest packing, and the translational and rotational dynamics of a confined guest, making it a useful benchmark for evaluating the transferability of ORION to molecular-crystal environments.

We first examine methane rotational dynamics using the second-order rotational autocorrelation function of the C-H bond vectors [Eq. (2)]. The rotational dynamics of methane confined in hydrate cages were characterized by the second-order rotational autocorrelation function of the C–H bond vectors,

$$C_2(\tau) = \langle P_2[\mathbf{u}_{m,i}(t) \cdot \mathbf{u}_{m,i}(t+\tau)] \rangle_{t,m,i} \qquad P_2(x) = \frac{1}{2}(3x^2 - 1) \qquad (2)$$

where $\mathbf{u}_{m,i}(t)$ is the unit vector of the $i$-th C–H bond in methane molecule $m$ at time $t$, and the averaging is performed over time origins, methane molecules, and the four C–H bonds.

In both models, the autocorrelation decays extremely rapidly and approaches zero on a sub-picosecond timescale, indicating that methane rotates very quickly inside the hydrate cages and reaches orientational equilibrium on an ultrafast timescale. Despite this common qualitative behavior, ORION predicts systematically slower orientational relaxation than TIP4P/Ice+GAFF. The fitted rotational correlation time increases from 0.0561 ps for TIP4P/Ice+GAFF to 0.0749 ps for ORION, corresponding to an increase of about 33.5%. Although the absolute timescales remain extremely short, this relative increase reveals a clear and systematic strengthening of guest confinement under ORION. Further support for this trend is provided by the distribution of methane off-center displacement relative to the cage center in Figure 8(c). In both models, the distributions are narrow and concentrated within sub-ångström displacements, showing that methane remains localized near the cage center rather than undergoing large off-center excursions. However, ORION predicts a more localized positional distribution, whereas TIP4P/Ice+GAFF allows broader translational excursions away from the cage center. A methane molecule that is more strongly confined near the cage center has access to fewer local configurations and is less able to undergo large-amplitude eccentric motion within the cavity; correspondingly, its orientational memory decays more slowly, leading to a longer rotational relaxation time. In this sense, the off-center displacement analysis is fully consistent with the slower decay of the P2-RACF under ORION. Although ORION does not qualitatively alter the ultrafast rotational character of encaged methane, it clearly predicts a somewhat stronger effective confinement of the guest molecule. The same picture is reflected in the O–C radial distribution function, as shown in Figure 8(d). Both models recover the characteristic shell structure of the hydrate crystal, indicating that the host framework remains intact. At the same time, ORION produces a slightly sharper first coordination peak, pointing to a somewhat more structured local host–guest packing environment. The difference is therefore not one of gross structural rearrangement, but of finer local organization.

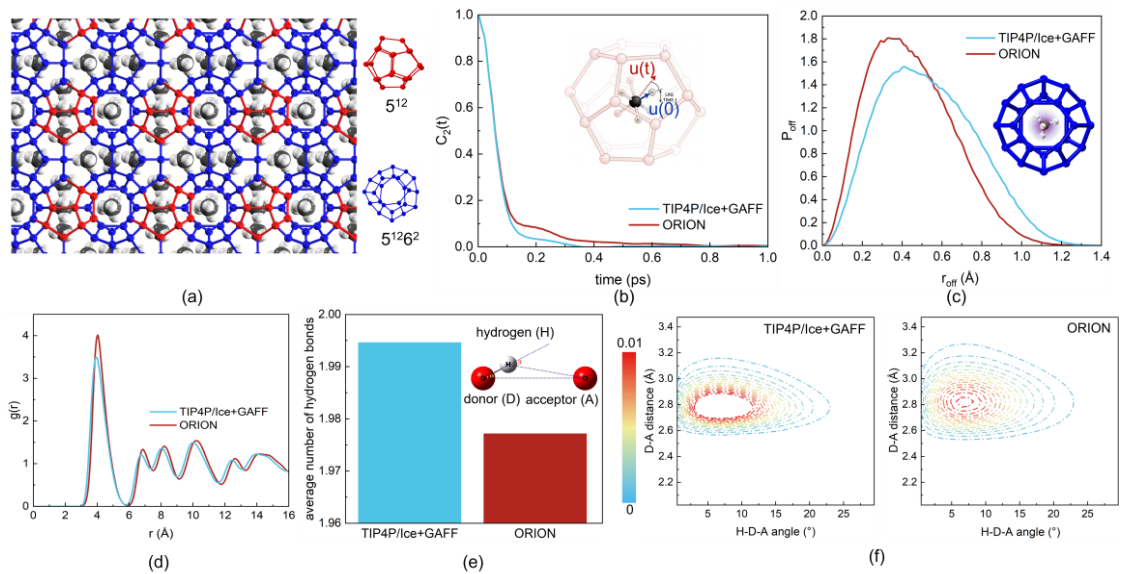

Figure 8. Structural and dynamical characterization of sI methane clathrate hydrate as a representative molecular-crystal benchmark. (a) Crystal structure of sI methane hydrate; (b) Methane P2 rotational autocorrelation function; (c) Methane off-center displacement distribution relative to the local cage center; (d) O(water)-C(CH$_4$) radial distribution function; (e) Average number of hydrogen bonds per water molecule; (f) Joint distribution of hydrogen-bond donor (D)–acceptor (A) distance and H–D–A angle

We next examine the water hydrogen-bond network, which provides the primary structural connectivity of the hydrate lattice. In this work, geometric standards are used to identify hydrogen bonds, where the distance between the proton donor (D) and acceptor (a) is less than 0.35 nm and the D-H ⋯ A angle is less than 30 °, indicating the presence of hydrogen bonds[47]. Figure 8(e) shows that ORION predicts a slightly smaller average number of hydrogen bonds per water molecule than TIP4P/Ice+GAFF, indicating a modest reduction in hydrogen-bond saturation. This is consistent with the joint distribution of D–A distance and H–D–A angle in Figure 8(f), where ORION exhibits a somewhat broader distribution, slightly longer donor–acceptor distances, and somewhat larger deviations from ideal linearity. These features suggest that the hydrate framework remains intact in ORION, but with a slightly softer and less overconstrained hydrogen-bond geometry than in the rigid TIP4P/Ice-based description.

## 3.4 Dynamics of Nucleic Acid and Protein

To further demonstrate the generality of the ORION force field beyond materials-oriented applications, we extend our simulations to biomolecular systems in aqueous solution. We consider two representative cases: a DNA–polycyclic aromatic hydrocarbon (PAH) complex in water and a solvated protein-ligand system, thereby probing the transferability of ORION to biomolecular environments.

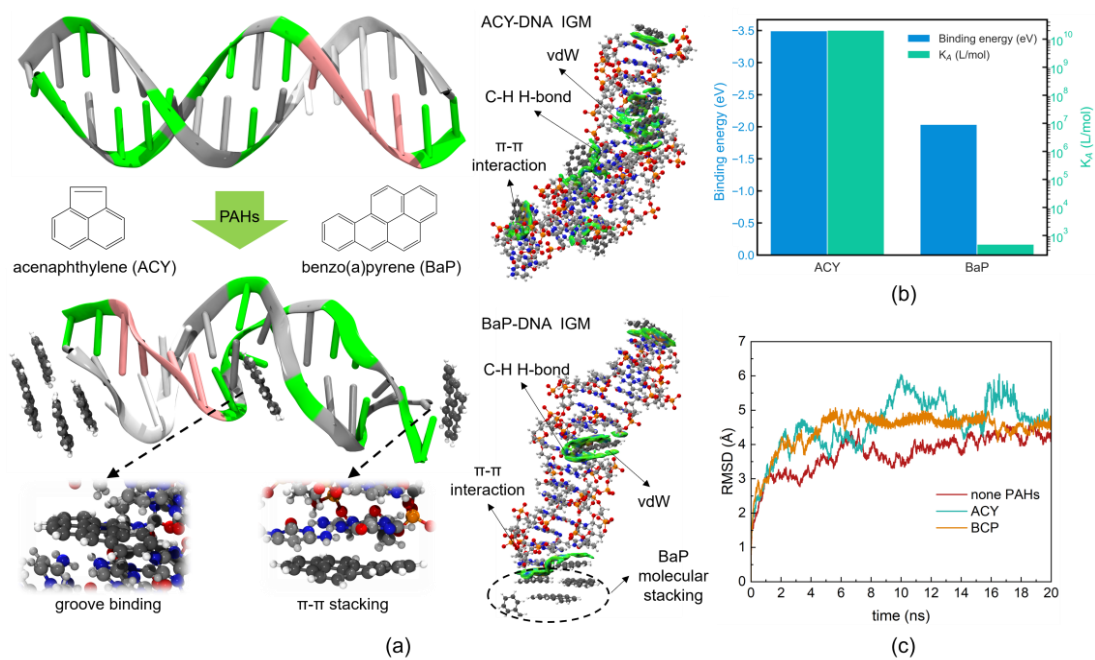

Figure 9. Non-bonding between polycyclic aromatic hydrocarbons and DNA. (a) Weak interaction between polycyclic aromatic hydrocarbons (PAHs) and DNA; (b) The binding energy between polycyclic aromatic hydrocarbons and DNA, as well as the fluorescence static quenching time constant ($K_A$); (c) Root mean square (RMSD) of DNA under different PAHs

PAHs are representative persistent organic pollutants whose genotoxicity is closely linked to their molecular recognition of DNA and the subsequent formation of PAH–DNA adducts[48-50]. To probe this recognition at the atomistic level, we constructed two PAH ensembles with comparable total mass but different aromatic sizes (11 acenaphthylene molecules, ACY; 7 benzo[a]pyrene molecules, BaP) and simulated their interactions with a commonly used short DNA duplex (GGCGGCGGCGTTTTGG) in explicit water. This DNA short chain is commonly used to study the binding of carbon particles to DNA[51, 52]. As shown in Figure 9(a), both PAHs preferentially adsorb in the DNA groove and insert between adjacent base pairs, indicating a groove-binding mode. This insertion locally disrupts base stacking and leads to an overall distortion of the duplex. Consistently, the DNA RMSD increases markedly in the presence of PAHs, with ACY inducing the largest deviation as shown in Figure 9(c). Using Multiwfn software[53-55], we performed an Independent Gradient Model (IGM) analysis to characterize the weak interactions at the PAH–DNA interface. The results indicate that PAH binding is primarily stabilized by nucleobase–PAH π-π interactions together with van der Waals (vdW) interactions and weak C–H hydrogen bonds. Among these contributions, the π-π interactions between PAHs and base pairs are particularly critical, directly promoting the formation of the groove-binding configuration. ORION yields a more favorable interaction energy for ACY binding to DNA than for BaP (−3.49 vs −2.03 eV). This stronger association is consistent with the larger DNA deformation observed for ACY, indicating a higher capability to perturb the duplex. This size trend also agrees with the fluorescence-quenching data compiled in Figure 9(b). In the cited experiments[56], the static quenching constant ($K_A$) is higher for the smaller PAH, supporting stronger ground-state complex formation with DNA. Mechanistically, larger PAHs such as BaP exhibit a greater tendency toward pi–pi self-stacking. This aggregation competes with PAH–DNA π-π contacts by sequestering aromatic surface area, thereby reducing the effective affinity for DNA. The equilibrated configurations reflect this behavior: BaP forms more pronounced stacked clusters, whereas ACY remains more available to engage nucleobases directly, which strengthens groove binding and enhances duplex distortion. In addition, along the molecular dynamics trajectories we observe intermittent proton-transfer events involving phosphate groups, surrounding water molecules, and nucleobase sites. Such behavior is generally inaccessible to conventional fixed-topology force fields (e.g., AMBER and CHARMM), where the bonding pattern is constrained and protons cannot relocate.

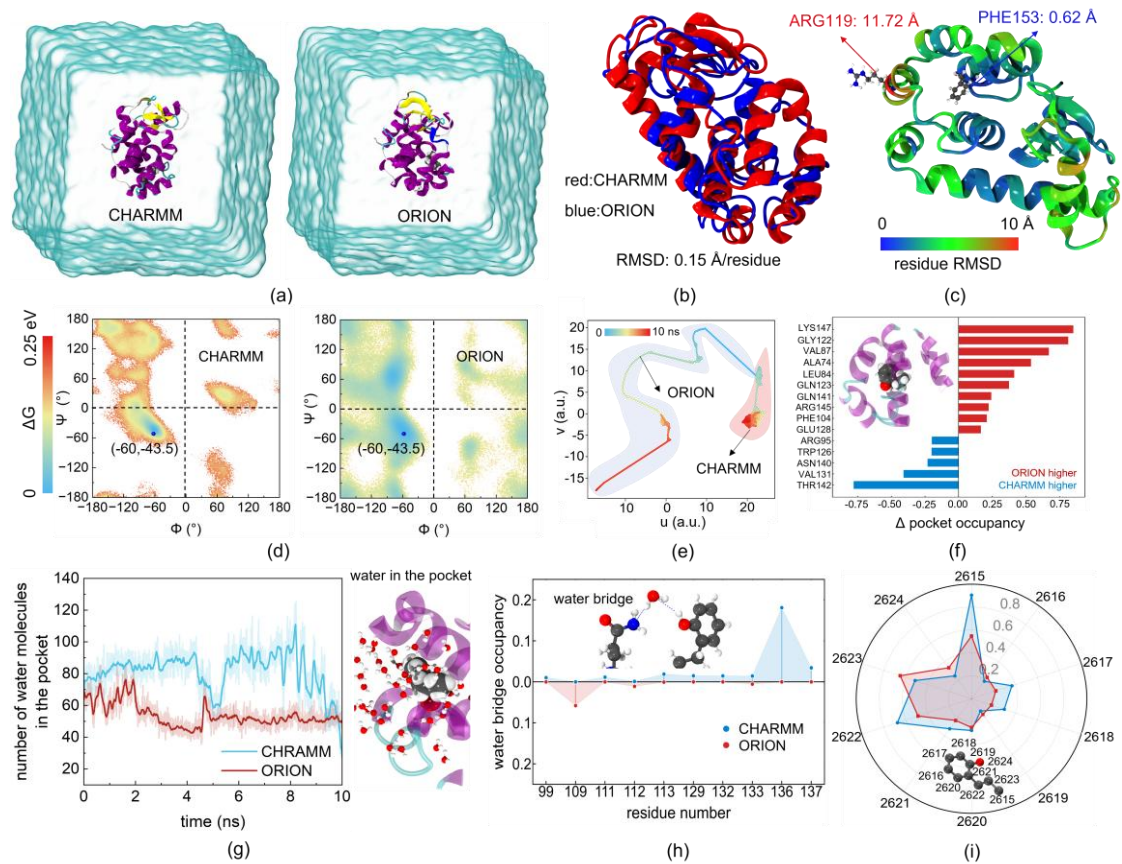

Figure 10. Comparison of ORION and CHARMM force fields for predicting protein-ligand configuration. (a) Final-frame protein conformations from CHARMM vs ORION simulations; (b) Structural superposition of CHARMM and ORION results with RMSD; (c) Per-residue RMSD distribution mapped onto the structure; (d) Free-energy changes along Ramachandran reaction coordinates ; (e) Comparison of ORION and CHARMM trajectories in UMA-reduced space; (f) Comparison of residue occupancy differences for the ligand in the binding pocket; (g) Time evolution of the number of water molecules in the pocket; (h) Comparison of water-bridge occupancy between the ligand and residues; (i) Comparison of ligand heavy-atom Root Mean Square Fluctuation (RMSF).

Proteins are another essential class of biomolecules, and atomistic MD simulations of protein–ligand complexes can quantify binding stability and reveal conformational changes, including transient pocket formation and water-mediated interactions relevant to drug discovery. As a representative benchmark, we simulated T4 lysozyme L99A/M102Q in complex with a ligand and compared ORION against CHARMM[57, 58]. The ligand molecule is 2-phenylphenol. The total simulation time is 20 ns, and the last 10ns are used for data analysis. The system was fully solvated in water, neutralized with $OH^-$ counterions to balance the net positive protein charge, and contained 40958 atoms in total.

As shown in Figure 10, for the T4 lysozyme L99A/M102Q–2-phenylphenol complex, ORION and CHARMM yield similar overall protein structures but differ in how binding microstates are populated. Figure 10(b) shows that structural superposition preserves the same dominant backbone scaffold in both force fields, with a difference of only 0.15 Å/residue, indicating that the overall protein structure described by ORION is highly reasonable. Consistently, the Ramachandran free-energy landscape exhibits an identical global minimum at $(\Phi,\Psi) \approx (-60°, -43.5°)$ in both simulations, demonstrating that ORION accurately captures the most stable backbone conformation. Beyond this shared minimum, ORION samples a broader region of low free energy, with additional basins and connected pathways in the $\Phi/\Psi$ space, which is also reflected by a more dispersed distribution in the UMA-reduced conformational space. These features indicate an expanded low-free-energy ensemble under ORION rather than a shift of the lowest-energy structure. The broader sampling can be attributed to differences in model formulation: CHARMM employs a fixed-topology empirical potential that restricts large-scale rearrangements, whereas ORION does not rely on predetermined topological constraints, allowing greater conformational accessibility under explicit solvent conditions. The differences between the two simulations are spatially localized. Residue-resolved heavy-atom RMSD analysis identifies ARG119 as the largest deviation point, with an RMSD difference of 11.72 Å, while most residues differ by less than 4 Å. In contrast, PHE153 shows a minimal RMSD of 0.62 Å, indicating nearly identical behavior. This pattern suggests that the divergence originates from pocket-coupled regions associated with local deformation and side-chain preorganization, rather than from global structural changes. These localized conformational differences directly translate into changes in the binding-pocket microenvironment. Compared with CHARMM, ORION stabilizes higher contact occupancies between the ligand and a larger number of pocket residues, indicating a more distributed protein–ligand interaction network. This redistribution is coupled to pocket hydration and water-mediated interactions. Under CHARMM, residue 136 exhibits the highest water-bridge occupancy (0.18), whereas under ORION the dominant water bridge shifts to residue 109 with a lower peak occupancy of 0.058, reflecting a reorganization of the water-bridge network topology. The altered interaction network has clear dynamical consequences for the ligand. The ligand heavy-atom RMSF is systematically smaller in the ORION simulations, indicating a more stabilized bound state with reduced flexibility. Overall, ORION preserves the same global minimum and overall fold as CHARMM while expanding the accessible low-free-energy microstate manifold and reshaping pocket contact and water-bridge networks. This ensemble-level description provides chemically relevant insight into binding heterogeneity, which is may important for drug-discovery applications where pocket flexibility and structured water play key roles in binding stability and selectivity.

## 4. Conclusion

In this work, we developed ORION, a universal machine-learning force field for organic systems, trained on a chemically complete CHONSP dataset constructed through an integrated top-down and bottom-up strategy. This framework enables simultaneous coverage of realistic macromolecular environments, reactive fragments, and weak intermolecular interactions, thereby substantially enhancing transferability across diverse organic and condensed-phase systems while retaining high computational efficiency. On the test set, ORION reproduces DFT-level atomic forces with substantially lower error than ReaxFF, while delivering more than two orders of magnitude acceleration in simulation speed.

Across representative applications, ORION consistently captures the essential atomistic chemistry of complex molecular systems. In combustion, it reproduces the key features of methane oxidation and resolves the oxygen-dependent combustion behavior of lignite. In carbon-material formation, it identifies the microscopic sequence of alkane pyrolysis from rapid cracking to aromatization, cluster growth, and graphitization. In systems dominated by weak interactions, ORION correctly distinguishes solvent effects in carbon nanotube dispersion, accurately describes hydrate-crystal host-guest behavior, and performs well in biomolecular applications including PAH-DNA recognition and protein-ligand binding. Taken together, these examples span bond breaking and formation, hydrogen bonding, van der Waals interactions, π-π stacking, and solvation effects, demonstrating that ORION provides a unified and transferable atomistic simulation platform for a broad range of problems in chemistry, materials science, and molecular biology.

Although the present model already covers a wide range of CHONSP systems, further extensions remain important. In particular, the explicit treatment of long-range electrostatics, charged species, and more diverse electronic states would broaden its applicability to more complex biomolecular and interfacial environments. We are therefore pursuing next-generation developments of ORION that incorporate these effects while preserving the efficiency and transferability established here. More broadly, this work highlights integrated top-down and bottom-up data construction as a practical route toward truly general machine-learning force fields for complex molecular systems.

## Acknowledgments

This research was supported by the National Natural Science Foundation of China (Nos. 42576237, 52206268, 52174255), the Shenzhen Science and Technology Program (No. JCYJ20250604144911016) and the Natural Science Foundation of Top Talent of SZTU (No. GDRC202314).

## Data availability:

Complete input and output files for the ORION model are freely available at https://gitlab.com/brucefan1983/nep-data.